\documentclass[reprint,aip,twocolumn,cha,showkeys,showpacs,superscriptaddress]{revtex4-1}

\usepackage{amsmath,amssymb,color,bm}
\usepackage{graphicx}
\renewcommand{\thetable}{\Roman{table}}
\renewcommand{\thesection}{\Roman{section}}
\usepackage[caption=false]{subfig}
\usepackage{booktabs}
\usepackage{color}
\usepackage{ulem} 
\usepackage{siunitx}

\renewcommand{\emph}{\textit}

\begin{document}
\title{Frequency control and coherent excitation transfer in a nanostring resonator network}
\date{\today}

\author{Matthias Pernpeintner}
\affiliation{Walther-Mei{\ss}ner-Institut, Bayerische Akademie der Wissenschaften, Walther-Mei\ss{}ner-Str.~8, D-85748 Garching, Germany}
\affiliation{Nanosystems Initiative Munich, Schellingstra{\ss}e 4, D-80799 M\"{u}nchen, Germany}
\affiliation{Physik-Department, Technische Universit\"{a}t M\"{u}nchen, James-Franck-Str.~1, D-85748 Garching, Germany}
\author{Philip Schmidt}
\affiliation{Walther-Mei{\ss}ner-Institut, Bayerische Akademie der Wissenschaften, Walther-Mei\ss{}ner-Str.~8, D-85748 Garching, Germany}
\affiliation{Nanosystems Initiative Munich, Schellingstra{\ss}e 4, D-80799 M\"{u}nchen, Germany}
\affiliation{Physik-Department, Technische Universit\"{a}t M\"{u}nchen, James-Franck-Str.~1, D-85748 Garching, Germany}
\author{Daniel Schwienbacher}
\affiliation{Walther-Mei{\ss}ner-Institut, Bayerische Akademie der Wissenschaften, Walther-Mei\ss{}ner-Str.~8, D-85748 Garching, Germany}
\affiliation{Nanosystems Initiative Munich, Schellingstra{\ss}e 4, D-80799 M\"{u}nchen, Germany}
\affiliation{Physik-Department, Technische Universit\"{a}t M\"{u}nchen, James-Franck-Str.~1, D-85748 Garching, Germany}
\author{Rudolf Gross}
\affiliation{Walther-Mei{\ss}ner-Institut, Bayerische Akademie der Wissenschaften, Walther-Mei\ss{}ner-Str.~8, D-85748 Garching, Germany}
\affiliation{Nanosystems Initiative Munich, Schellingstra{\ss}e 4, D-80799 M\"{u}nchen, Germany}
\affiliation{Physik-Department, Technische Universit\"{a}t M\"{u}nchen, James-Franck-Str.~1, D-85748 Garching, Germany}
\author{Hans Huebl}
\email{huebl@wmi.badw.de}
\affiliation{Walther-Mei{\ss}ner-Institut, Bayerische Akademie der Wissenschaften, Walther-Mei\ss{}ner-Str.~8, D-85748 Garching, Germany}
\affiliation{Nanosystems Initiative Munich, Schellingstra{\ss}e 4, D-80799 M\"{u}nchen, Germany}
\affiliation{Physik-Department, Technische Universit\"{a}t M\"{u}nchen, James-Franck-Str.~1, D-85748 Garching, Germany}

\begin{abstract}
Coupling, synchronization, and non-linear dynamics of resonator modes are omnipresent in nature \cite{pikovsky_synchronization_2003} and highly relevant for a multitude of applications ranging from lasers to Josephson arrays and spin torque oscillators. Nanomechanical resonators are ideal candidates to study these effects on a fundamental level \cite{shim_synchronized_2007,karabalin_nonlinear_2009,heinrich_photon_2010,holmes_synchronization_2012,bagheri_photonic_2013,matheny_phase_2014} and to realize all-mechanical platforms for information processing \cite{hatanaka_phonon_2013,rips_quantum_2013,faust_coherent_2013} and storage\cite{oconnell_quantum_2010,palomaki_coherent_2013}. For larger resonator networks, however, this requires the ability to tune the mode frequencies selectively and to operate the resonators in the strong coupling regime. Here, we present a proof-of-principle realization of a resonator network consisting of two high-quality nanostring resonators, coupled mechanically by a shared support. First, we demonstrate that we can control the fundamental mode frequencies of both nanostrings independently by a strong drive tone resonant with one of the higher harmonics of the network, rendering local control gates redundant \cite{unterreithmeier_universal_2009}. The tuning mechanism relies on an effective increase of the pre-stress in a highly excited nanostring, known as geometric nonlinearity. Using this selective frequency control of the individual nanomechanical resonators, we investigate the coherent dynamics of the resonator network, which is a classical model system showing several of the characteristic features of strongly coupled quantum systems. In particular, we demonstrate mode splitting, classical Rabi oscillations, as well as adiabatic and diabatic transitions between the coupled states representing the classical analog of Landau-Zener tunneling. Therefore, this coupling and tuning concept opens the path to a selective phonon transfer between two spatially separated mechanical resonators.
\end{abstract}

\maketitle

Information processing and storage in a network of mechanical resonators \cite{hatanaka_phonon_2013,hatanaka_phonon_2014,mahboob_multimode_2014} requires the control and targeted transfer of phonons between different resonators. In particular, the ability to independently address and tune the individual mechanical resonators is a key requirement for multi-resonator networks. 
Here, we employ a simple scheme for tuning the eigenfrequency of multiple nanostring resonators on a chip individually via a globally applied drive. Based on the geometric nonlinearity of tensile stressed nanostring resonators, we make use of a higher-mode excitation to modify the effective stress in the nanostring without any local control gates. 
Tuning the mechanical mode frequency in this way does not affect the readout signal-to-noise ratio or contribute additional damping mechanisms. This allows the operation of nanostring resonators with constant and high $Q$-factors over a tuning range of tens of $\mathrm{kHz}$.

In our experiment, we control the effective stress in the nanostring by strongly driving an auxiliary mode (here we use the second-order mode, $n=2$) at frequency $\Omega_{\mathrm{aux}}$ (see also Refs.~\citenum{karabalin_nonlinear_2009,westra_nonlinear_2010}). For large excitation amplitudes, the effective length of the nanostring increases, which corresponds to an effective increase of the tensile stress and a shift of the nanostring's eigenfrequency. We operate the auxillary mode in the Duffing regime, where its amplitude is tunable via the drive frequency \cite{nayfeh_nonlinear_1979}. Thus, the resonance frequency of the fundamental mode is controlled by varying the drive frequency of one of the higher harmonics of the same nanostring. 
A quantitative analysis (see Supplementary Information) shows that the shift of the fundamental frequency $\Delta\Omega_{m}$ is related to the auxillary drive frequency $\Omega_{\mathrm{aux}}$ (independent of the auxillary mode index $n$) by
\begin{equation}\label{eq:FundVsCtrlFreq}
	\frac{\Delta\Omega_{m}}{\Omega_{m}}=\frac{2}{3}\frac{\Omega_{\mathrm{aux}}-\Omega_{m,n}}{\Omega_{m,n}}.
\end{equation}
Here, $\Omega_{m}$ and $\Omega_{m,n}$ are the eigenfrequencies of the fundamental and auxillary mode of resonator $m$.

\begin{figure}[tbh]
	\centering
	\includegraphics[scale=1]{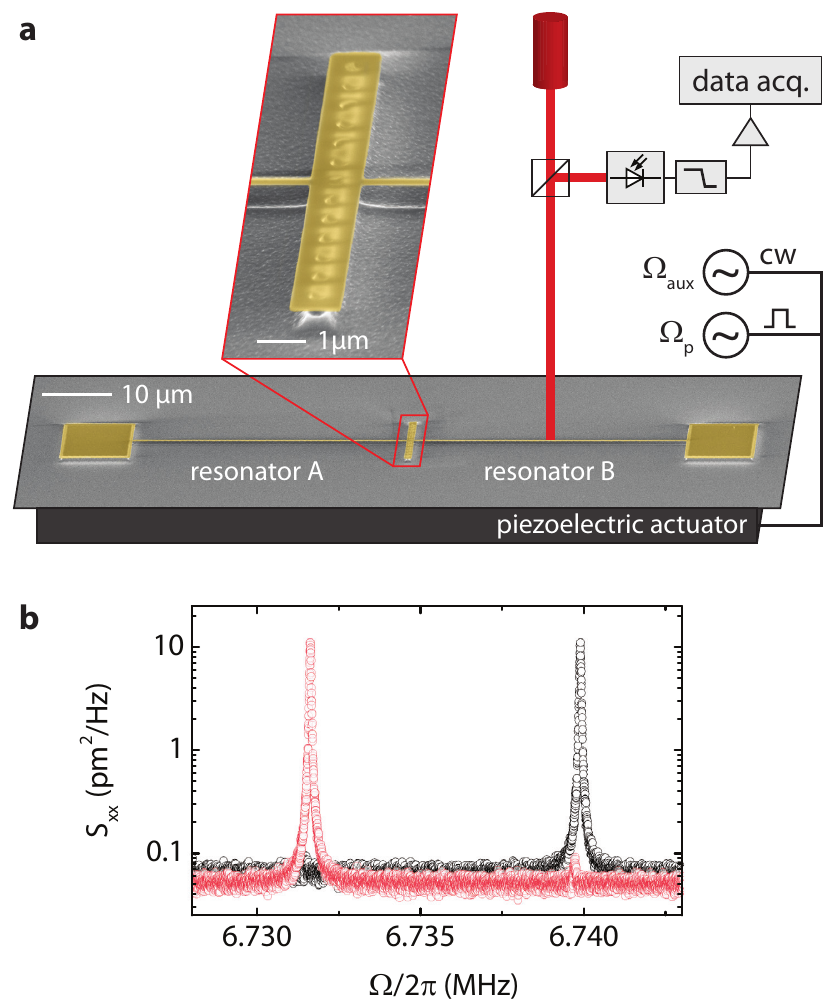}
	\caption{\textit{Experimental setup and thermal motion spectra.} \textbf{a.} False-colored scanning electron microscope image of the sample (tilt angle: \SI{60}{\degree}) and schematic illustration of the experimental setup. The two nanostrings are mechanically coupled by a shared support which is partially suspended. The whole chip is mounted on top of a piezoelectric actuator, which is driven with two radio-frequency tones: a strong continuous auxillary drive (which is used to tune the eigenfrequency of nanostring A) and a pulsed drive (used to excite nanostring A or B). For readout of the mechanical motion, a laser ($\lambda=\SI{633}{nm}$) is focused either on nanostring A or B. The reflected laser light is guided to a photodetector, filtered, amplified and detected with a spectrum analyzer or a digitizer card. \textbf{b.} Thermal motion spectra obtained by focusing the laser on nanostring A (red circles) or nanostring B (black circles) without an auxillary drive applied. 
	}
	\label{fig:Assembly-Setup}
\end{figure}

Besides the frequency tunability of individual resonators, the ability to coherently transfer information between resonators is a key requirement for the implementation of mechanical resonator networks. Here, we investigate two $l=\SI{40}{\micro m}$ long, tensile stressed silicon nitride nanostring resonators, coupled mechanically by a shared support (see Fig.~1a). The two resonators show slightly different eigenfrequencies $\Omega_{\mathrm{A}}^0/2\pi=\SI{6.732}{MHz}$ and $\Omega_{\mathrm{B}}^0/2\pi=\SI{6.740}{MHz}$ with a (FWHM) linewidth of $\Gamma_{\mathrm{A,B}}/2\pi\approx \SI{42}{Hz}$ at room temperature, corresponding to a quality factor of $\num{1.6e5}$. 
The sample is mounted in vacuum to prevent air damping of the nanostring's motion. Both nanostrings can be excited and frequency-controlled with a global external driving force provided by a piezoelectric actuator. For the readout of the mechanical displacement, we employ optical laser interferometry as sketched in Fig.~\ref{fig:Assembly-Setup}a (cf.~e.\,g.~Ref.~\citenum{vogel_optically_2003}). By focusing the laser spot on one of the two nanostrings, we selectively measure its displacement as a function of time, giving access to the dynamic evolution of the coupled system as well as its thermal motion spectrum. All experiments shown here are performed at room temperature.

Figure~\ref{fig:Assembly-Setup}b displays the measured thermal motion spectra when probing nanostring A (red circles) and B (black circles) without applying any drive. Here, the frequency detuning of the two nanostrings is much larger than the coupling rate, resulting in two well separated resonances at frequencies $\Omega_{\mathrm{A}}^0$ and $\Omega_{\mathrm{B}}^0$. These resonances correspond to the fundamental out-of-plane modes of nanostring A and B.

Next, we show that the nanostrings can be tuned in resonance using the above described tuning mechanism. To this end, we employ a strong auxillary drive via the piezoelectric actuator, exciting the second-order mode (\textit{auxillary mode}) of nanostring A.

\begin{figure}[tbh]
	\centering
	\includegraphics[scale=1]{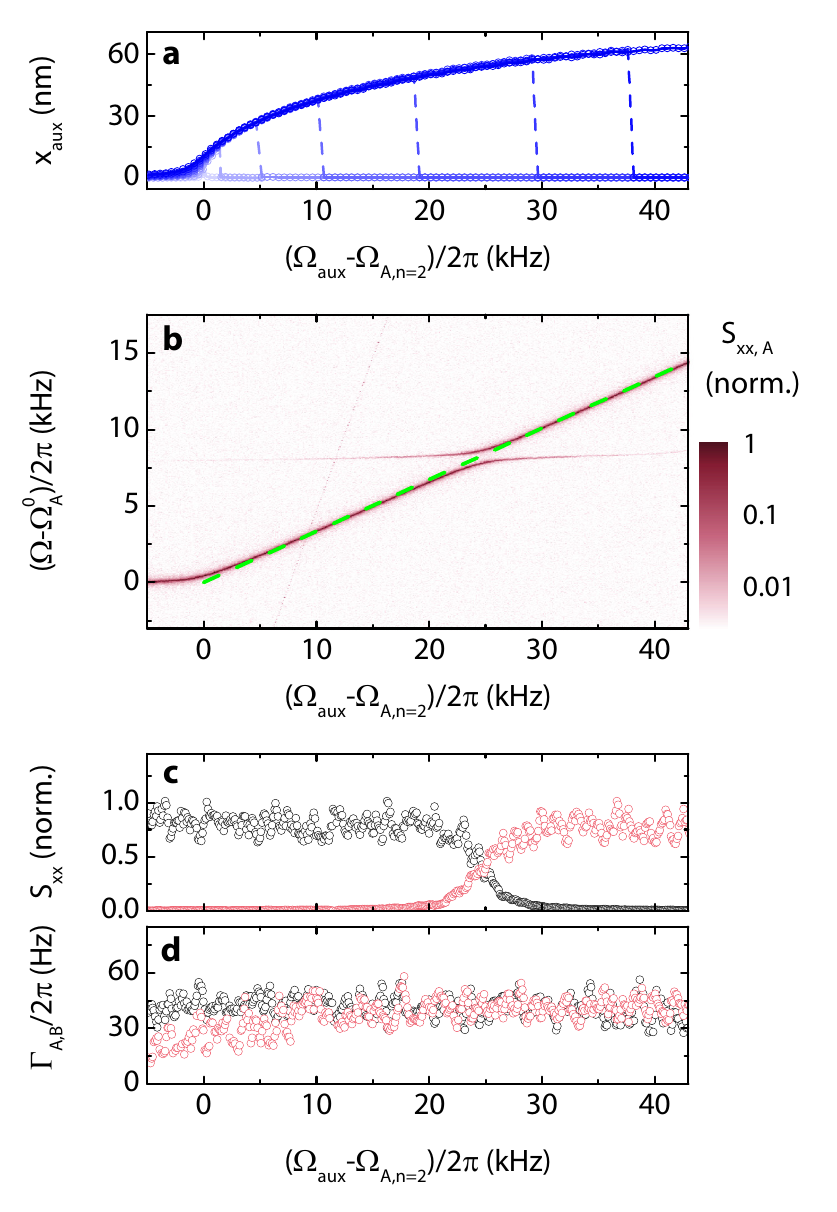}
	\caption{\textit{Frequency tuning and mode splitting.} \textbf{a.} Amplitude spectrum of the second-order mode (\textit{auxillary mode}) of nanostring A for the drive amplitudes $\SI{30}{mV_{rms}}$, $\SI{125}{mV_{rms}}$, $\SI{300}{mV_{rms}}$, $\SI{400}{mV_{rms}}$, $\SI{530}{mV_{rms}}$, $\SI{710}{mV_{rms}}$, $\SI{940}{mV_{rms}}$, and $\SI{1.3}{V_{rms}}$ shown as open circles colored from light to dark blue. \textbf{b.} Thermal motion spectrum of nanostring A as a function of the auxillary drive frequency. The auxillary drive amplitude is $V_{\mathrm{aux}}=\SI{3.1}{V_{rms}}$. The dashed green line indicates the expected eigenfrequency shift given by Eq.~(\ref{eq:FundVsCtrlFreq}). \textbf{c-d.} Amplitude (\textbf{c}) and linewidth (\textbf{d}) of the thermal motion spectra shown in \textbf{b}. The red circles indicate the upper mode, the black circles denote the lower mode.}
	\label{fig:Assembly-ModeTuning}
\end{figure}

Figure~\ref{fig:Assembly-ModeTuning}a shows the amplitude spectrum of the auxillary mode for a strong coherent drive. As expected for a Duffing oscillator with positive cubic nonlinearity, we observe an increase of the effective resonance frequency for strong drive amplitudes. The displacement $x_{\mathrm{aux}}$ of the auxillary mode is related to the drive frequency via $x_{\mathrm{aux}}\propto \sqrt{\Omega_{\mathrm{aux}}-\Omega_{\mathrm{A},n=2}}$ (Ref.~\citenum{nayfeh_nonlinear_1979}), where $\Omega_{\mathrm{A},n=2}$ is the eigenfrequency of the auxillary mode. We use this well-known behavior to tune the eigenfrequency of the fundamental mode. To this end, we measure the thermal motion spectrum of the fundamental out-of-plane mode of nanostring A as a function of the auxillary drive frequency $\Omega_{\mathrm{aux}}$, as shown in Fig.~\ref{fig:Assembly-ModeTuning}b. We use a constant auxillary drive amplitude and gradually increase the frequency $\Omega_{\mathrm{aux}}$, starting slightly below $\Omega_{\mathrm{A},n=2}$. While for $\Omega_{\mathrm{aux}}<\Omega_{\mathrm{A},n=2}$, the auxillary mode amplitude is small and hence the fundamental mode frequency is nearly constant, the tuning effect is clearly visible for $\Omega_{\mathrm{aux}}-\Omega_{\mathrm{A},n=2}\gg \Gamma_{\mathrm{A},n=2}$ (where $\Gamma_{\mathrm{A},n=2}/2\pi\approx \SI{220}{Hz}$ is the linewidth of the auxillary mode). Here, the fundamental mode frequency quantitatively follows our model depicted as green dashed line in Fig.~\ref{fig:Assembly-ModeTuning}b. Figure~\ref{fig:Assembly-ModeTuning}c,d shows that the linewidth and the signal amplitude of the fundamental mode are not affected by the auxillary drive, enabling a high signal-to-noise ratio and maintaining an excellent and constant mechanical quality factor over the full frequency tuning range. 

Around $\Omega_{\mathrm{aux}}-\Omega_{\mathrm{A},n=2}\approx 2\pi\times\SI{25}{kHz}$, we observe an avoided crossing of the two fundamental out-of-plane modes of nanostring A and B, with a mode splitting $g/2\pi=\SI{830}{Hz}$. Even though the coupling rate $g$ is small compared to the eigenfrequencies of the modes, the nanostrings are deeply in the strong coupling regime due to their small damping rates $\Gamma_{\mathrm{A,B}} \ll g$. This allows investigating the coherent dynamics of two spatially separated, mechanically coupled nanostring resonators including the exchange and controlled transfer of excitations. Theoretically, our system can be described as a fully classical two-level system with negative linear coupling between the amplitudes of the two nanostring resonators (see Supplementary Information).

\begin{figure*}[tbh!]
	\centering
	\includegraphics[scale=1]{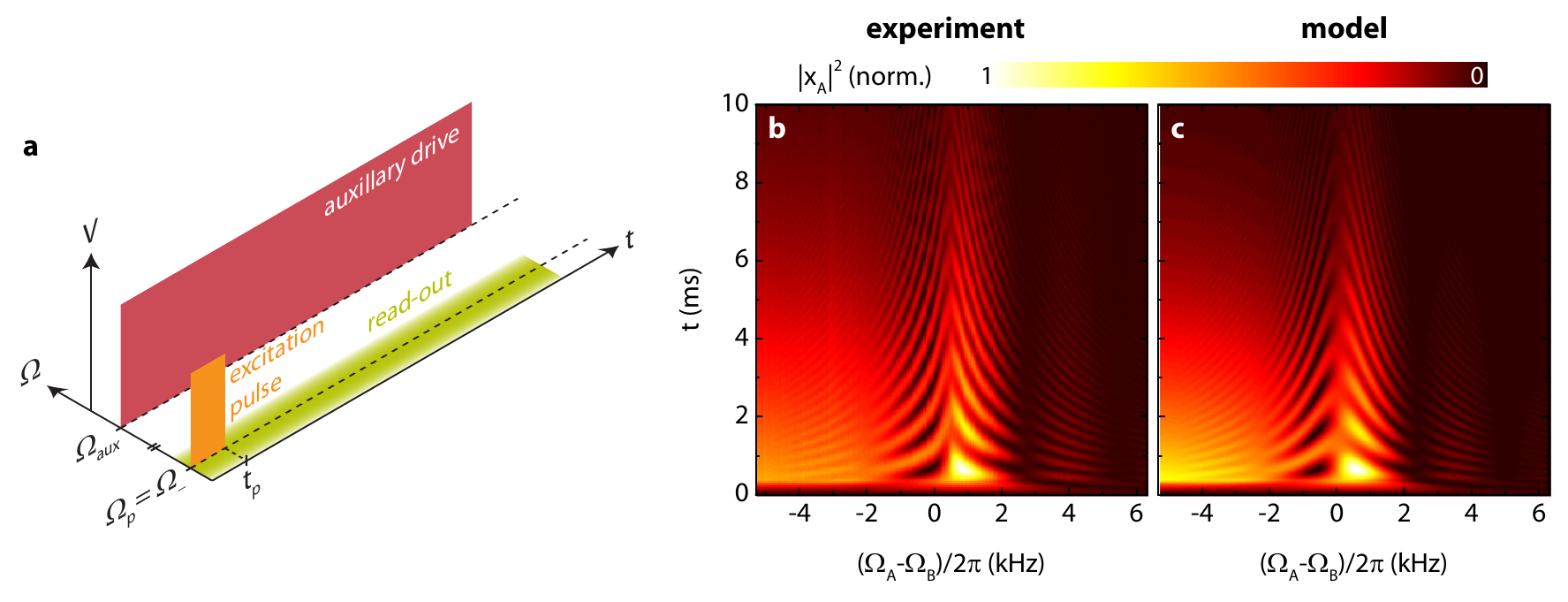}
	\caption{\textit{Rabi oscillations.} \textbf{a.} Measurement scheme: After a short excitation pulse (length $t_{\mathrm{p}}=\SI{400}{\micro s}$, amplitude $V_{\mathrm{p}}=\SI{0.22}{V_{rms}}$) at the lower mode's frequency $\Omega_{-}$, the displacement of nanostring A is measured as a function of time. A constant drive tone (drive voltage $V_{\mathrm{aux}}=\SI{3.1}{V_{rms}}$) at the auxillary frequency $\Omega_{\mathrm{aux}}$ is used to tune the eigenfrequency $\Omega_{\mathrm{A}}$ of nanostring A. \textbf{b.} Squared displacement amplitude of the fundamental mode of nanostring A as a function of time $t$ and eigenfrequency $\Omega_{\mathrm{A}}$. \textbf{c.} Model results, obtained by solving the equations of motion as detailed in the Supplementary Information. Experimental observations and the model calculations show quantitative agreement.}
	\label{fig:Assembly-Rabi}
\end{figure*}

We start with the coherent oscillatory exchange of excitations between the fundamental out-of-plane modes of nanostring A and B, the classical analog of Rabi oscillations in a quantum two-level system (see also Refs.~\citenum{faust_coherent_2013,okamoto_coherent_2013}). We initialize the system by exciting the lower mode resonantly with a short pulse of duration $t_{\mathrm{p}}$ and amplitude $V_{\mathrm{p}}$ as sketched in Fig.~\ref{fig:Assembly-Rabi}a. In a single-shot experiment, we measure the displacement of nanostring A as a function of time, followed by digital down-conversion and low-pass filtering. This allows to study the time evolution of the energy stored in this resonator, which is proportional to the square of its motional amplitude. 

In Fig.~\ref{fig:Assembly-Rabi}b, the squared oscillation amplitude of nanostring A is plotted versus time as a function of the eigenfrequency $\Omega_{\mathrm{A}}$. We can distinguish two regimes: For $\Omega_{\mathrm{A}}-\Omega_{\mathrm{B}}\ll-g$, we find a simple exponential decay of the excitation of nanostring A. When both modes are close to resonance, we observe a coherent oscillation of the energy stored in the mechanical motion between both nanostrings. On resonance, this oscillation frequency is given by the coupling constant $g$. For higher $\Omega_{\mathrm{A}}-\Omega_{\mathrm{B}}$, the signal vanishes as the lower mode now corresponds to nanostring B; thus the excitation pulse of the initialization sequence is mainly localized in nanostring B.
This qualitative picture is quantitatively corroborated in Fig.~\ref{fig:Assembly-Rabi}c showing a numerical simulation of the dynamics of this classical two-level system (for details see Supplementary Information).

\begin{figure*}[tbh]
	\centering
	\includegraphics[scale=1]{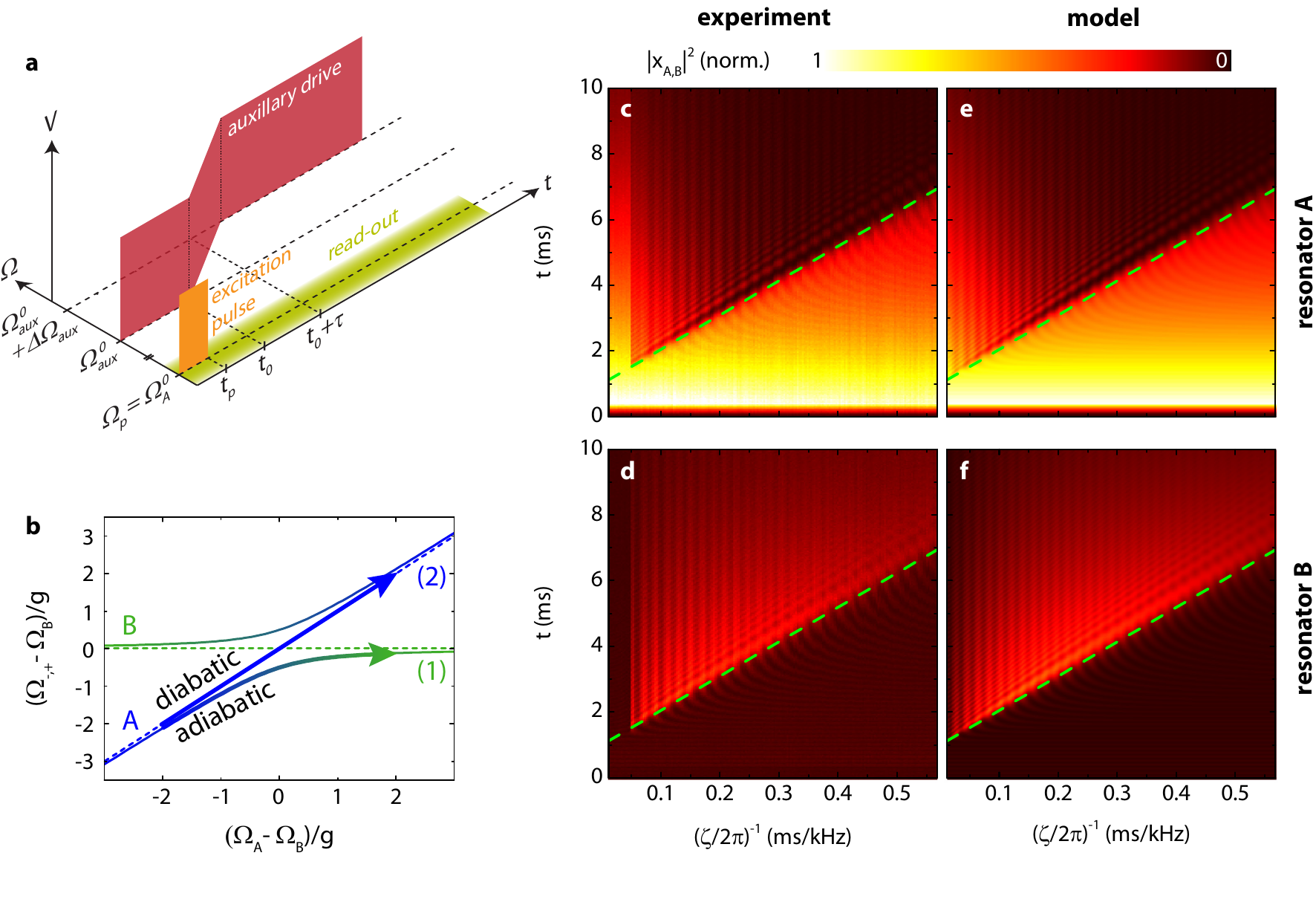}
	\caption{\textit{Landau-Zener transitions.} \textbf{a.} Measurement scheme: After a short excitation pulse (length $t_{\mathrm{p}}=\SI{400}{\micro s}$, amplitude $V_{\mathrm{p}}=\SI{0.22}{V_{rms}}$) with frequency $\Omega_{\mathrm{A}}^0$, the displacement of the fundamental mode of nanostring A (or B) is measured as a function of time. An auxillary drive tone (drive voltage $V_{\mathrm{aux}}=\SI{4.2}{V_{rms}}$) with frequency $\Omega_{\mathrm{aux}}$ is used to ramp the eigenfrequency $\Omega_{\mathrm{A}}$ of nanostring A from $\Omega_{\mathrm{A}}^0<\Omega_{\mathrm{B}}$ to $\Omega_{\mathrm{A}}^0+\Delta\Omega_{\mathrm{A}}>\Omega_{\mathrm{B}}$ with ramp rate $\zeta=\Delta\Omega_{\mathrm{A}}/\tau$. \textbf{b.} Schematic illustration of adiabatic and diabatic transitions: For slow tuning of nanostring A (\textit{adiabatic transition}), the system follows the lower branch of the mode splitting and ends up in state (1), i.\,e.~the excitation of nanostring A is transferred to nanostring B. For fast frequency ramps, the system passes the mode crossing diabatically without energy transfer between the nanostrings (final state: (2)). \textbf{c,d.} Squared displacement amplitude of the fundamental mode of nanostring A (\textbf{c}) and B (\textbf{d}) as a function of time $t$ and frequency ramp rate $\zeta$. \textbf{e,f.} Model results, obtained by solving the equations of motion as detailed in the Supplementary Information. The dashed green lines in \textbf{c}-\textbf{f} indicate the time $t_{\mathrm{s}}$ at which nanostring A and B are resonant, $\Omega_{\mathrm{A}}=\Omega_{\mathrm{B}}$. Experimental observations and the model calculations show quantitative agreement.  }
	\label{fig:Assembly-LandauZener}
\end{figure*}

For information processing with phonons, targeted phonon transfer between coupled mechanical resonators is required\cite{seitner_classical_2016,fu_classical_2016}. We perform a classical Landau-Zener-type experiment\cite{faust_nonadiabatic_2012}, demonstrating this ability in our mechanically coupled nanostrings. Using a short excitation pulse, we initialize nanostring A and subsequently increase its eigenfrequency with constant ramp rate $\zeta$ from $\Omega_{\mathrm{A}}^0$ to $\Omega_{\mathrm{A}}^0+\Delta\Omega_{\mathrm{A}}$ (with $\Omega_{\mathrm{A}}^0<\Omega_{\mathrm{B}}<\Omega_{\mathrm{A}}^0+\Delta\Omega_{\mathrm{A}}$), as sketched in Fig.~\ref{fig:Assembly-LandauZener}a. 

In analogy to quantum mechanical Landau-Zener transitions, we expect the final state of the system to depend on the frequency ramp rate. For slow tuning of nanostring A, $\zeta\ll \zeta_{\mathrm{tr}} = \pi g^2/(2\ln{2})$ (Ref.~\citenum{faust_nonadiabatic_2012}), the system adiabatically follows the lower branch of the mode splitting as sketched in Fig.~\ref{fig:Assembly-LandauZener}b and ends up in state (1) corresponding to an excitation of nanostring B. Thus, the energy stored in nanostring A is fully transferred to nanostring B. For fast frequency ramps, however, the diabatic behavior dominates and the system passes the transition region without energy transfer.

Figure~\ref{fig:Assembly-LandauZener}c,e shows the time evolution of the energy stored in nanostring A and B as a function of the frequency ramp rate $\zeta$. To take these data, we have performed the experiment twice while probing nanostring A or B, respectively. 

For very fast frequency ramps ($\zeta > \zeta_{\mathrm{c}} \approx \SI{20}{kHz/ms}$), the bandwidth of the auxillary mode limits the speed of the Duffing response and in particular prevents sustained large amplitude oscillations required for the frequency tuning mechanism. Therefore, the fundamental mode becomes independent of the auxillary drive and we observe a simple exponential decay of the energy stored in nanostring A. This sets the upper limit of the applicable frequency tuning rate $\zeta$. To circumvent this limitation in future experiments, a higher damping rate of the auxillary mode is desirable. This can be realized either by choosing a higher mode index for the auxillary mode\cite{unterreithmeier_damping_2010} or by mode engineering (see e.\,g.~Refs.~\citenum{ghadimi_dissipation_2016,tsaturyan_ultra-coherent_2016}).
 
For frequency ramp rates below $\zeta_{\mathrm{c}}$, the fundamental mode's eigenfrequency follows the auxillary drive frequency and reaches the resonance condition $\Omega_{\mathrm{A}}=\Omega_{\mathrm{B}}$ at time $t_{\mathrm{s}}(\zeta)$, indicated by the dashed green line in Fig.~\ref{fig:Assembly-LandauZener}c-f. Around $t=t_{\mathrm{s}}$, the nanostrings interact with each other, with the final state depending on the frequency ramp rate $\zeta$. 

In the case of small ramp rates ($\zeta\lesssim \zeta_{\mathrm{tr}}\approx \SI{10}{kHz/ms}$), we observe a distinct excitation transfer from nanostring A to B. For $\zeta\approx \zeta_{\mathrm{tr}}$, the excitation is only partially transferred to nanostring B, i.\,e.~the system is in the transition region between adiabatic and diabatic behavior. For $\zeta>\zeta_{\mathrm{tr}}$, the excitation mainly maintains in resonator A as expected. Besides, the described overall behavior is superimposed with interference effects stemming from coherent Rabi-like oscillations between nanostring A and B around $t=t_{\mathrm{s}}$. Yet, all these features are quantitatively explained by numerically solving the equations of motion with the system parameters determined above (Fig.~\ref{fig:Assembly-LandauZener}d,f). Note that there are no free fit parameters in this model (for details see Supplementary Information).

In conclusion, we demonstrate coherent dynamics in a system of two spatially separated, high-$Q$ nanostring resonators coupled by a shared support. For frequency tuning of the resonators, we use a globally applied coherent driving force. This technique, which is based on the geometric nonlinearity of tensile stressed nanostrings, allows to address individual resonators on a single chip independently without the use of local control gates. Hence this work presents a promising approach for the efficient manipulation of multiple coupled nanostring resonators and opens the path towards targeted transfer of excitations in phonon networks.

\pagebreak
\widetext
\vspace{1cm}
\begin{center}
	\textbf{\large Supplementary Information: Frequency control and coherent excitation transfer in a nanostring resonator network}
\end{center}

\renewcommand{\thetable}{\Roman{table}}
\renewcommand{\thesection}{\Roman{section}}
\renewcommand*{\citenumfont}[1]{S#1}
\renewcommand*{\bibnumfmt}[1]{[S#1] }

\makeatletter
\makeatother

\renewcommand{\thesection}{\Alph{section}}
\renewcommand{\thesubsection}{\alph{subsection}}
\renewcommand{\thefigure}{S\arabic{figure}}
\renewcommand{\thetable}{S\arabic{table}}
\renewcommand{\theequation}{S\arabic{equation}}
\renewcommand{\refname}{Additional References}

\setcounter{equation}{0}


\section{Experimental details}
\subsection{Sample fabrication}
The sample was fabricated from a single-crystalline silicon substrate, covered with a $\SI{90}{nm}$ thin tensile-stressed silicon nitride film. Using electron-beam lithography, aluminium evaporation and lift-off, the $l=\SI{40}{\micro m}$ long and $\SI{120}{nm}$ wide nanostrings and their supports were defined. The pattern was transferred to the silicon nitride film by anisotropic reactive ion etching (RIE) with an Ar/SF$_6$ process. Using a subsequent isotropic RIE step with SF$_6$ only, the nanostrings were released. In this step, the etching time was chosen in a way that the shared support remained weakly connected to the substrate. In the final cleaning procedure, the aluminium hard mask was removed.
	
\subsection{Experimental setup} 
The sample was mounted on a piezoelectric actuator, which was connected to two radio-frequency sources. The first source was continuously driving the auxillary mode (in order to tune the resonance frequency of nanostring A), the second source was used to excite one of the coupled modes with short drive pulses. To avoid air damping of the mechanical motion, the sample was operated in vacuum ($p<\SI{1e-4}{mbar}$). For readout of the nanostring motion, we used a laser interferometer with a HeNe laser ($\SI{2.2}{mW}$, $\lambda=\SI{633}{nm}$) focused on one of the two nanostring resonators. Direct optical access to the sample via an integrated microscope allowed to precisely focus the laser on the desired position and thus detect the motion of both nanoresonators individually. The reflected laser light was guided to a biased silicon photodetector. The measured photovoltage was lowpass-filtered, amplified and recorded with a spectrum analyzer or a digitizer card (sampling rate: $\SI{100}{MHz}$). 
	
\subsection{Data post-processing and amplitude calibration} 
The time-domain data were digitally downconverted, downsampled and lowpass-filtered (cutoff frequency: $\SI{25}{kHz}$). For calibration of the auxillary mode displacement amplitude (see Fig.~\ref{fig:Assembly-ModeTuning}a), we measured the response spectrum of this mode for increasing drive amplitude and fitted the shift of the effective resonance frequency to the expected backbone curve (similar to Ref.~\citenum{pernpeintner_circuit_2014}). The relevant material parameters (density and Young's modulus of the silicon nitride film) have been taken from Ref.~\citenum{pernpeintner_circuit_2014}. To determine the thermal motion amplitudes of nanostring A and B (Fig.~\ref{fig:Assembly-Setup}b), we used a calibration based on the equipartition theorem\cite{saulson_thermal_1990}.

\section{Duffing nonlinearity and effective elongation of a highly excited nanostring}
In this section, we relate the Duffing nonlinearity of a tensile stressed nanostring resonator to material and geometry parameters. Besides, we derive an expression for the effective (i\,.e.~time-averaged) elongation of the nanostring when one of its higher-order modes is excited. These relations are used in Sec.~\ref{SI_sec:TuningMechanism} for a quantitative description of the eigenfrequency tuning mechanism.

For a highly tensile-stressed nanostring, the displacement $x$ as a function of the coordinate along the nanostring axis, $\eta$, can be well approximated by\cite{cleland_foundations_2003}  
\begin{equation*}
	x(\eta)=x_{0,n}\sin(n\pi\eta/l),\quad 0\leq\eta\leq l,
\end{equation*}
where $n$ is the mode index, $x_{0,n}$ is the displacement at the center of the nanostring and $l$ denotes the length of the string.

A non-vanishing displacement $x_{0,n}$ induces an elongation of the string according to
\begin{equation}\label{SI_eq:ldash}
	l'=\int_{0}^{l} \sqrt{1+\left(\frac{\mathrm{d}}{\mathrm{d}\eta}x(\eta)\right)^2}\,\mathrm{d} \eta\approx \int_{0}^{l} \left(1+\frac{1}{2}\left(\frac{\mathrm{d}}{\mathrm{d}\eta}x(\eta)\right)^2\right)\mathrm{d}\eta = l\left(1+\frac{n^2x_{0,n}^2\pi^2}{4l^2}\right).
\end{equation}
This elongation increases the tensile stress $\sigma$ along the string axis,
\begin{equation*}
	\sigma=\sigma_0+E \frac{l'-l}{l}=\sigma_0+\frac{n^2x_{0,n}^2\pi^2 E}{4l^2},
\end{equation*}
where $\sigma_0$ and $E$ denote the pre-stress and Young's modulus of the nanostring.

Substituting this into the equation of motion of a tensile stressed nanostring\cite{weaver_vibration_1990},
\begin{equation}\label{SI_eq:eomHTS}
	\ddot{x}(t)+\left(\frac{n\pi}{l}\right)^2\frac{\sigma}{\rho}x(t)=0,
\end{equation}
where $\rho$ is the material density, yields
\begin{equation*}
	\ddot{x}(t)+\left(\frac{n\pi}{l}\right)^2\frac{\sigma_0}{\rho}x(t)+\frac{n^4\pi^4 E}{4l^4\rho}x^3(t)=0.
\end{equation*}
Thus, the elongation of the displaced nanostring leads to an additional term in the equation of motion which is proportional to $x^3(t)$. This \textit{geometric nonlinearity} can be described by defining the nonlinearity parameter $\alpha_n$ as
\begin{equation}\label{SI_eq:alphan_definition}
	\alpha_n:=\frac{n^4\pi^4 E}{4l^4\rho}.
\end{equation}
With this, the Equation of motion (\ref{SI_eq:eomHTS}) takes the standard form of a Duffing oscillator\cite{nayfeh_nonlinear_1979,unterreithmeier_coherent_2009}
\begin{equation*}
	\ddot{x}(t)+\Omega_n^2 x(t)+\alpha_n x^3(t)=0
\end{equation*}
with the \textit{Duffing parameter} $\alpha_n$ and the resonance frequency $\Omega_n=n\pi/l\sqrt{\sigma_0/\rho}$.

For tuning the eigenfrequency of the nanostring by exciting a higher-order auxillary mode, the effective (i.\,e.~time-averaged) elongation of the string, caused by the auxillary mode, is of interest. Time-averaging Eq.~(\ref{SI_eq:ldash}) yields: 
\begin{equation}\label{SI_eq:effectiveElongation}
	\Delta l:=\langle l'-l \rangle=\frac{1}{2} \frac{n^2x_{0,n}^2\pi^2}{4l}
\end{equation}
Here, we have used the harmonic time-dependence $x(t)\propto \exp(i\Omega_n t)$.

We will use Eq.~(\ref{SI_eq:effectiveElongation}) in the following to quantitatively describe the eigenfrequency tuning mechanism.

\section{Quantitative description of the eigenfrequency tuning mechanism}\label{SI_sec:TuningMechanism}
To tune the eigenfrequency of individual nanostrings in a resonator network, we use the geometric nonlinearity of tensile stressed nanostring resonators. We drive a higher-order mode to a high amplitude state using a globally applied strong coherent driving force. This leads to an effective elongation of the string, which increases the tensile stress and thus the eigenfrequency of the fundamental mode. In this section, we derive a quantitative description of this mode tuning effect and relate the eigenfrequency shift to the auxillary drive frequency. 

The resonance frequency of the fundamental mode of a highly tensile stressed nanostring resonator (called \textit{resonator $m$} in the following) can be well approximated by
\begin{equation*}
	\Omega_{m}=\frac{\pi}{l}\sqrt{\frac{\sigma}{\rho}}
\end{equation*}
with the length $l$ of the nanostring, the density $\rho$ and the pre-stress $\sigma$ (Ref.~\citenum{verbridge_high_2006}). A static elongation $\Delta l$ of the nanostring modifies the pre-stress according to
\begin{equation*}
	\sigma'=\sigma_0+\Delta\sigma \quad\text{with}\quad \Delta\sigma=E\frac{\Delta l}{l},
\end{equation*}
where $E$ is Young's modulus. The resonance frequency is then given by
\begin{equation*}
	\Omega'_{m}=\frac{\pi}{l}\sqrt{\frac{\sigma'}{\rho}}\approx \Omega_{m} \left(1+\frac{\Delta\sigma}{2\sigma_0}\right).
\end{equation*}

To generate such an effective static elongation $\Delta l$, we apply a strong drive to excite the $n$\textsuperscript{th}-order mode (\textit{auxillary mode}) of the nanostring. For a given amplitude of the auxillary mode oscillation, $x_{\mathrm{aux}}\equiv x_{0,n}$, the effective, i.\,e.~time-averaged, elongation is (see Eq.~\ref{SI_eq:effectiveElongation})
\begin{equation*}
	\Delta l=\frac{1}{2}\frac{n^2 x_{0,n}^2\pi^2}{4l}.
\end{equation*}
With this, the relative eigenfrequency change is
\begin{equation}\label{SI_eq:RelFreqChange}
	\frac{\Delta\Omega_{m}}{\Omega_{m}}:=\frac{\Omega'_{m}-\Omega_{m}}{\Omega_{m}}=\frac{x_{0,n}^2}{4\sigma_0}\frac{n^2\pi^2E}{4l^2}.
\end{equation}

Next, we relate the auxillary mode amplitude $x_{0,n}$ to the auxillary drive frequency $\Omega_{\mathrm{aux}}$. To this end, we start with the amplitude spectrum of a Duffing oscillator, which is given by the implicit equation
\begin{equation}\label{SI_eq:DuffingAmplSpec}
	\left[\Gamma_n^2+4(\Omega_{\mathrm{aux}}-\Omega_{m,n}-\frac{3}{8}\frac{\alpha_n}{\Omega_{m,n}}x_0^2)^2\right]x_0^2=\frac{F_0^2}{m^2\Omega_{m,n}^2}.
\end{equation}
Here, $\Omega_{m,n}$ and $\Gamma_n$ are the eigenfrequency and (FWHM) linewidth of the $n$\textsuperscript{th}-order mode and $\Omega_{\mathrm{aux}}$ is the excitation frequency. $F_0$, $m$ and $x_{0,n}$ denote the driving force, the effective mass of the nanostring and the amplitude of the excited auxillary mode. The Duffing nonlinearity $\alpha_n$ is related to material parameters via (see Eq.~(\ref{SI_eq:alphan_definition}))
\begin{equation}\label{SI_eq:def_alphaN}
	\alpha_n=\frac{n^4\pi^4 E}{4l^4\rho}.
\end{equation}

In case of high amplitudes ($x_{0,n}^3\gg 4F_0/(3\alpha_n m)$), Eq.~(\ref{SI_eq:DuffingAmplSpec}) can be simplified to
\begin{equation*}
	x_{0,n}^2=\frac{8}{3}\frac{\Omega_{m,n}}{\alpha_n}(\Omega_{\mathrm{aux}}-\Omega_{m,n}),
\end{equation*}
which holds for $x_{0,n}\leq F_0/(\Gamma_n m\Omega_{m,n})$. Note that this relation is independent of the driving force $F_0$, thus the amplitude $x_{0,n}$ only depends on the excitation frequency $\Omega_{\mathrm{aux}}$.

Substituting this into Eq.~(\ref{SI_eq:RelFreqChange}) and using Eq.~(\ref{SI_eq:def_alphaN}), we obtain
\begin{equation}\label{SI_eq:FundVsCtrlFreq}
	\frac{\Delta\Omega_{m}}{\Omega_{m}}=\frac{2}{3}\frac{\Omega_{\mathrm{aux}}-\Omega_{m,n}}{\Omega_{m,n}}.
\end{equation}
Remarkably, the prefactor $2/3$ is independent of the mode index $n$ of the auxillary mode.

Equation~(\ref{SI_eq:FundVsCtrlFreq}) illustrates that the fundamental mode frequency -- or any other mode's eigenfrequency -- can be controlled by a strong external drive with tunable frequency $\Omega_{\mathrm{aux}}$. Please note that the bistability of the highly excited auxillary mode requires an initialization of the auxillary mode at $\Omega_{\mathrm{aux}}\approx\Omega_{m,n}$ and a subsequent continuous sweep of the auxillary drive frequency to the target value $\Omega_{\mathrm{aux}}$.

\section{Coupled nanostrings: Equations of motion and mode splitting}
To describe our system of two coupled nanostring resonators, we follow the derivation given in Refs.~\citenum{novotny_strong_2010,seitner_classical_2016}. We start with two harmonic resonators with (undisturbed) resonance frequencies $\tilde{\Omega}_{\mathrm{A}}=\sqrt{k_{\mathrm{A}}/m_{\mathrm{A}}}$ and $\tilde{\Omega}_{\mathrm{B}}=\sqrt{k_{\mathrm{B}}/m_{\mathrm{B}}}$. Assuming a linear coupling between the resonators, the equations of motion read
\begin{align}\label{SI_eq:eom_modesplitting}
	\begin{split} 
		m_{\mathrm{A}}\ddot{x}_{\mathrm{A}}+k_{\mathrm{A}}x_{\mathrm{A}} &= k_{\mathrm{g}}(x_{\mathrm{B}}-x_{\mathrm{A}}) \\
		m_{\mathrm{B}}\ddot{x}_{\mathrm{B}}+k_{\mathrm{B}}x_{\mathrm{B}} &= k_{\mathrm{g}}(x_{\mathrm{A}}-x_{\mathrm{B}}).
	\end{split} 
\end{align} 

Due to the geometry of the shared support, the coupling parameter $k_{\mathrm{g}}$ is negative. This means that a positive displacement of nanostring A results in a negative force on nanostring B, leading to a negative displacement of nanostring B.

We search for solutions of the form $x_m(t)=x_m^0\exp(i\Omega_{\pm}t)$ ($m={\mathrm{A,B}}$). Substituting this ansatz into Eqs.~(\ref{SI_eq:eom_modesplitting}), we obtain the resonance frequencies of the hybrid states
\begin{equation*}
	\Omega_{\pm}^2=\frac{1}{2}\left[\Omega_{\mathrm{A}}^2+\Omega_{\mathrm{B}}^2\pm\sqrt{\left(\Omega_{\mathrm{A}}^2-\Omega_{\mathrm{B}}^2\right)^2+4g^2\Omega_{\mathrm{A}}\Omega_{\mathrm{B}}}\right]
\end{equation*}
with $\Omega_m^2=(k_m+k_{\mathrm{g}})/m_m$ and $g^2=k_{\mathrm{g}}^2/(m_{\mathrm{A}}m_{\mathrm{B}}\Omega_{\mathrm{A}}\Omega_{\mathrm{B}})$.

As we are dealing with two nearly identical nanostring resonators, we set $m_{\mathrm{A}}=m_{\mathrm{B}}\equiv m$, $k_{\mathrm{B}}\equiv k$ and $k_{\mathrm{A}}=k+\Delta k$ with $\Delta k\ll k$. In this case, $g^2\approx k_{\mathrm{g}}^2/(m k)$. In our experiment, $\Delta k$ can be tuned via an external control parameter. For $|\Delta k|\gg k_{\mathrm{g}}$, the resonators A and B are nearly uncoupled and the mode frequencies $\Omega_{\pm}$ correspond to the bare mechanical frequencies $\Omega_{\mathrm{A}}$ and $\Omega_{\mathrm{B}}$. If the modes are tuned close to resonance ($|\Delta k|\lesssim k_{\mathrm{g}}$), the coupling leads to a hybridization of the modes and a modification of the mechanical eigenfrequencies. For $\Delta k=0$, we obtain a mode splitting $\Omega_+-\Omega_- = g$.

Using the measured thermal motion spectra shown in Fig.~2b (main text), we find $g/2\pi=\SI{830}{Hz}$, $\Omega_{\mathrm{A}}^0/2\pi=\SI{6.732}{MHz}$ and $\Omega_{\mathrm{B}}^0/2\pi=\SI{6.740}{MHz}$. This leads to the following sample parameters:
\begin{align}\label{SI_eq:sample_params}
	\begin{split} 
		m &= \SI{0.67}{pg} \\
		k &= \SI{1.20}{kg/s^2} \\
		k_{\mathrm{g}} &= \SI{-1.5e-4}{kg/s^2}
	\end{split} 
\end{align}

\section{Rabi oscillations}
In this section, we model the coherent oscillation of excitations in the coupled nanostring system (classical Rabi oscillations). Adding a linear damping term and an external coherent drive to the Equations of motion~(\ref{SI_eq:eom_modesplitting}), we obtain
\begin{align}\label{SI_eq:eom_Rabi}
	\begin{split} 
		m_{\mathrm{A}}\ddot{x}_{\mathrm{A}}+m_{\mathrm{A}}\Gamma_{\mathrm{A}}\dot{x}_{\mathrm{A}}+k_{\mathrm{A}}x_{\mathrm{A}} &= k_{\mathrm{g}}(x_{\mathrm{B}}-x_{\mathrm{A}})+F_{\mathrm{drive}} \\
		m_{\mathrm{B}}\ddot{x}_{\mathrm{B}}+m_{\mathrm{B}}\Gamma_{\mathrm{B}}\dot{x}_{\mathrm{B}}+k_{\mathrm{B}}x_{\mathrm{B}} &= k_{\mathrm{g}}(x_{\mathrm{A}}-x_{\mathrm{B}})+F_{\mathrm{drive}}.
	\end{split} 
\end{align} 
In the following, we set $\Gamma_{\mathrm{B}}=\Gamma_{\mathrm{A}}\equiv \Gamma$ and use $m_{\mathrm{A}}=m_{\mathrm{B}}\equiv m$, $k_{\mathrm{B}}\equiv k$ and $k_{\mathrm{A}}=k+\Delta k$ as detailed above.

For the investigation of Rabi oscillations, we excite the lower mode of the coupled nanostring system with a short coherent radio-frequency pulse, $F_{\mathrm{drive}}(t)=F_0\exp(i\Omega_{\mathrm{p}} t)\quad (0<t<t_{\mathrm{p}})$. The drive frequency corresponds to the lower mode's eigenfrequency, $\Omega_{\mathrm{p}}=\Omega_{-}$. Note that in the experiment, we drive the nanostring motion with a piezoelectric actuator, so the driving force acts on both nanostrings equally.  

We are looking for solutions of the form $x_m(t)=x_0 c_m(t)\exp(i\Omega_{\mathrm{p}} t)$ with $|c_{\mathrm{A}}|^2+|c_{\mathrm{B}}|^2=1$. Substituting this ansatz into Eqs.~(\ref{SI_eq:eom_Rabi}), we obtain
\begin{align}\label{SI_eq:eom_Rabi_Cs}
	\begin{split} 
		\ddot{c}_{\mathrm{A}}+(2i\Omega_{\mathrm{p}}+\Gamma)\dot{c}_{\mathrm{A}}+(\Omega_{\mathrm{A}}^2-\Omega_{\mathrm{p}}^2+i\Omega_{\mathrm{p}}\Gamma)c_{\mathrm{A}} &= \frac{k_{\mathrm{g}}}{m}c_{\mathrm{B}}+\frac{F_0}{m x_0}\theta(t_{\mathrm{p}}-t) \\
		\ddot{c}_{\mathrm{B}}+(2i\Omega_{\mathrm{p}}+\Gamma)\dot{c}_{\mathrm{B}}+(\Omega_{\mathrm{B}}^2-\Omega_{\mathrm{p}}^2+i\Omega_{\mathrm{p}}\Gamma)c_{\mathrm{B}} &= \frac{k_{\mathrm{g}}}{m}c_{\mathrm{A}}+\frac{F_0}{m x_0}\theta(t_{\mathrm{p}}-t).
	\end{split} 
\end{align} 
Here, $\theta(t)$ denotes the Heaviside step function.

For weak coupling $|k_{\mathrm{g}}| \ll k$, the oscillation of the coefficients $c_m(t)$ is much slower than the oscillatory term $\exp(i\Omega_{\mathrm{p}} t)$. Therefore, we can neglect the second derivatives in Eqs.~(\ref{SI_eq:eom_Rabi_Cs}) and obtain
\begin{align}
	\begin{split} 
		(2i\Omega_{\mathrm{p}}+\Gamma)\dot{c}_{\mathrm{A}}+(\Omega_{\mathrm{A}}^2-\Omega_{\mathrm{p}}^2+i\Omega_{\mathrm{p}}\Gamma)c_{\mathrm{A}} &= \frac{k_{\mathrm{g}}}{m}c_{\mathrm{B}}+\frac{F_0}{m x_0}\theta(t_{\mathrm{p}}-t) \nonumber\\
		(2i\Omega_{\mathrm{p}}+\Gamma)\dot{c}_{\mathrm{B}}+(\Omega_{\mathrm{B}}^2-\Omega_{\mathrm{p}}^2+i\Omega_{\mathrm{p}}\Gamma)c_{\mathrm{B}} &= \frac{k_{\mathrm{g}}}{m}c_{\mathrm{A}}+\frac{F_0}{m x_0}\theta(t_{\mathrm{p}}-t). \nonumber
	\end{split} 
\end{align} 

Using this set of equations and the experimentally determined parameters (see Eqs.~(\ref{SI_eq:sample_params})), we can quantitatively reproduce the experimental findings as Fig.~3 (main text) illustrates.

\section{Landau-Zener transitions}
In the Landau-Zener-type experiment, we start with the two mechanical modes far off resonance, $\Omega_{\mathrm{A}}(t=0)-\Omega_{\mathrm{B}}\ll -g$. We excite nanostring A (corresponding to the lower mode) with a short coherent pulse (length $t_{\mathrm{p}}$, frequency $\Omega_{\mathrm{p}}=\Omega_{\mathrm{A}}^0$) and then ramp its eigenfrequency upwards beyond $\Omega_{\mathrm{B}}$:
\begin{equation*}
	\Omega_{\mathrm{A}}(t)=
	\begin{cases}
		\Omega_{\mathrm{A}}^0 & t < t_0 \\
		\Omega_{\mathrm{A}}^0+\zeta (t-t_0) & t_0 \leq t < t_0+\tau \\
		\Omega_{\mathrm{A}}^0+\Delta\Omega_{\mathrm{A}} & t \geq t_0+\tau
	\end{cases}
\end{equation*}
with $\zeta=\Delta\Omega_{\mathrm{A}}/\tau$. We search for solutions of the form $x_m(t)=x_0 c_m(t)\exp(i\Omega_{\mathrm{A}}(t) t)$ ($m=\mathrm{A,B}$).

Due to the time-dependence of $\Omega_{\mathrm{A}}$, the equations of motion for $c_{\mathrm{A}}$ and $c_{\mathrm{B}}$ take the more complex form
\begin{align}\label{SI_eq:eom_LZ}
	\begin{split} 
		G(t)\dot{c}_{\mathrm{A}}+(F(t)+\Omega_{\mathrm{A}}^2(t))c_{\mathrm{A}} &= \frac{k_{\mathrm{g}}}{m}c_{\mathrm{B}}+\frac{F_0}{m x_0}\theta(t_{\mathrm{p}}-t) \\
		G(t)\dot{c}_{\mathrm{B}}+(F(t)+\Omega_{\mathrm{B}}^2)c_{\mathrm{B}} &=\frac{k_{\mathrm{g}}}{m}c_{\mathrm{A}}+\frac{F_0}{m x_0}\theta(t_{\mathrm{p}}-t),
	\end{split} 
\end{align} 
where we have defined
\begin{align}
	\begin{split} 
		F(t) &= (i\dot{\Omega}_{\mathrm{A}} t+i \Omega_{\mathrm{A}})^2+2i\dot{\Omega}_{\mathrm{A}}+\Gamma(i\dot{\Omega}_{\mathrm{A}} t+i \Omega_{\mathrm{A}}) \nonumber\\
		G(t) &= 2i(\dot{\Omega}_{\mathrm{A}}t+\Omega_{\mathrm{A}})+\Gamma. \nonumber
	\end{split} 
\end{align} 
Again, we have neglected the second derivatives of $c_{\mathrm{A}}$ and $c_{\mathrm{B}}$ as justified above.

Numerical solution of Eqs.~(\ref{SI_eq:eom_LZ}) yields Fig.~4d,f (main text), which is in very good agreement with the experimentally determined behavior of the coupled nanostring system.


\section*{References}
\begin{thebibliography}{10}
	\expandafter\ifx\csname url\endcsname\relax
	\def\url#1{\texttt{#1}}\fi
	\expandafter\ifx\csname urlprefix\endcsname\relax\def\urlprefix{URL }\fi
	\providecommand{\bibinfo}[2]{#2}
	\providecommand{\eprint}[2][]{\url{#2}}
	
	\bibitem{pikovsky_synchronization_2003}
	\bibinfo{author}{Pikovsky, A.}, \bibinfo{author}{Rosenblum, M.} \&
	\bibinfo{author}{Kurths, J.}
	\newblock \emph{\bibinfo{title}{Synchronization: A Universal Concept in
			Nonlinear Sciences}} (\bibinfo{publisher}{Cambridge University Press},
	\bibinfo{year}{2003}).
	
	\bibitem{shim_synchronized_2007}
	\bibinfo{author}{Shim, S.-B.}, \bibinfo{author}{Imboden, M.} \&
	\bibinfo{author}{Mohanty, P.}
	\newblock \bibinfo{title}{Synchronized {Oscillation} in {Coupled}
		{Nanomechanical} {Oscillators}}.
	\newblock \emph{\bibinfo{journal}{Science}} \textbf{\bibinfo{volume}{316}},
	\bibinfo{pages}{95} (\bibinfo{year}{2007}).
	
	\bibitem{karabalin_nonlinear_2009}
	\bibinfo{author}{Karabalin, R.~B.}, \bibinfo{author}{Cross, M.~C.} \&
	\bibinfo{author}{Roukes, M.~L.}
	\newblock \bibinfo{title}{Nonlinear dynamics and chaos in two coupled
		nanomechanical resonators}.
	\newblock \emph{\bibinfo{journal}{Physical Review B}}
	\textbf{\bibinfo{volume}{79}}, \bibinfo{pages}{165309}
	(\bibinfo{year}{2009}).
	
	\bibitem{heinrich_photon_2010}
	\bibinfo{author}{Heinrich, G.}, \bibinfo{author}{Harris, J. G.~E.} \&
	\bibinfo{author}{Marquardt, F.}
	\newblock \bibinfo{title}{Photon shuttle: Landau-zener-st\"uckelberg dynamics
		in an optomechanical system}.
	\newblock \emph{\bibinfo{journal}{Physical Review A}}
	\textbf{\bibinfo{volume}{81}}, \bibinfo{pages}{011801}
	(\bibinfo{year}{2010}).
	
	\bibitem{holmes_synchronization_2012}
	\bibinfo{author}{Holmes, C.~A.}, \bibinfo{author}{Meaney, C.~P.} \&
	\bibinfo{author}{Milburn, G.~J.}
	\newblock \bibinfo{title}{Synchronization of many nanomechanical resonators
		coupled via a common cavity field}.
	\newblock \emph{\bibinfo{journal}{Physical Review E}}
	\textbf{\bibinfo{volume}{85}}, \bibinfo{pages}{066203}
	(\bibinfo{year}{2012}).
	
	\bibitem{bagheri_photonic_2013}
	\bibinfo{author}{Bagheri, M.}, \bibinfo{author}{Poot, M.},
	\bibinfo{author}{Fan, L.}, \bibinfo{author}{Marquardt, F.} \&
	\bibinfo{author}{Tang, H.~X.}
	\newblock \bibinfo{title}{Photonic {Cavity} {Synchronization} of
		{Nanomechanical} {Oscillators}}.
	\newblock \emph{\bibinfo{journal}{Physical Review Letters}}
	\textbf{\bibinfo{volume}{111}}, \bibinfo{pages}{213902}
	(\bibinfo{year}{2013}).
	
	\bibitem{matheny_phase_2014}
	\bibinfo{author}{Matheny, M.~H.} \emph{et~al.}
	\newblock \bibinfo{title}{Phase {Synchronization} of {Two} {Anharmonic}
		{Nanomechanical} {Oscillators}}.
	\newblock \emph{\bibinfo{journal}{Physical Review Letters}}
	\textbf{\bibinfo{volume}{112}}, \bibinfo{pages}{014101}
	(\bibinfo{year}{2014}).
	
	\bibitem{hatanaka_phonon_2013}
	\bibinfo{author}{Hatanaka, D.}, \bibinfo{author}{Mahboob, I.},
	\bibinfo{author}{Onomitsu, K.} \& \bibinfo{author}{Yamaguchi, H.}
	\newblock \bibinfo{title}{A phonon transistor in an electromechanical resonator
		array}.
	\newblock \emph{\bibinfo{journal}{Applied Physics Letters}}
	\textbf{\bibinfo{volume}{102}}, \bibinfo{pages}{213102}
	(\bibinfo{year}{2013}).
	
	\bibitem{rips_quantum_2013}
	\bibinfo{author}{Rips, S.} \& \bibinfo{author}{Hartmann, M.~J.}
	\newblock \bibinfo{title}{Quantum {Information} {Processing} with
		{Nanomechanical} {Qubits}}.
	\newblock \emph{\bibinfo{journal}{Physical Review Letters}}
	\textbf{\bibinfo{volume}{110}}, \bibinfo{pages}{120503}
	(\bibinfo{year}{2013}).
	
	\bibitem{faust_coherent_2013}
	\bibinfo{author}{Faust, T.}, \bibinfo{author}{Rieger, J.},
	\bibinfo{author}{Seitner, M.~J.}, \bibinfo{author}{Kotthaus, J.~P.} \&
	\bibinfo{author}{Weig, E.~M.}
	\newblock \bibinfo{title}{Coherent control of a classical nanomechanical
		two-level system}.
	\newblock \emph{\bibinfo{journal}{Nature Physics}}
	\textbf{\bibinfo{volume}{9}}, \bibinfo{pages}{485} (\bibinfo{year}{2013}).
	
	\bibitem{oconnell_quantum_2010}
	\bibinfo{author}{O'Connell, A.~D.} \emph{et~al.}
	\newblock \bibinfo{title}{Quantum ground state and single-phonon control of a
		mechanical resonator}.
	\newblock \emph{\bibinfo{journal}{Nature}} \textbf{\bibinfo{volume}{464}},
	\bibinfo{pages}{697} (\bibinfo{year}{2010}).
	
	\bibitem{palomaki_coherent_2013}
	\bibinfo{author}{Palomaki, T.~A.}, \bibinfo{author}{Harlow, J.~W.},
	\bibinfo{author}{Teufel, J.~D.}, \bibinfo{author}{Simmonds, R.~W.} \&
	\bibinfo{author}{Lehnert, K.~W.}
	\newblock \bibinfo{title}{Coherent state transfer between itinerant microwave
		fields and a mechanical oscillator}.
	\newblock \emph{\bibinfo{journal}{Nature}} \textbf{\bibinfo{volume}{495}},
	\bibinfo{pages}{210} (\bibinfo{year}{2013}).
	
	\bibitem{unterreithmeier_universal_2009}
	\bibinfo{author}{Unterreithmeier, Q.~P.}, \bibinfo{author}{Weig, E.~M.} \&
	\bibinfo{author}{Kotthaus, J.~P.}
	\newblock \bibinfo{title}{Universal transduction scheme for nanomechanical
		systems based on dielectric forces}.
	\newblock \emph{\bibinfo{journal}{Nature}} \textbf{\bibinfo{volume}{458}},
	\bibinfo{pages}{1001} (\bibinfo{year}{2009}).
	
	\bibitem{hatanaka_phonon_2014}
	\bibinfo{author}{Hatanaka, D.}, \bibinfo{author}{Mahboob, I.},
	\bibinfo{author}{Onomitsu, K.} \& \bibinfo{author}{Yamaguchi, H.}
	\newblock \bibinfo{title}{Phonon waveguides for electromechanical circuits}.
	\newblock \emph{\bibinfo{journal}{Nature Nanotechnology}}
	\textbf{\bibinfo{volume}{9}}, \bibinfo{pages}{520} (\bibinfo{year}{2014}).
	
	\bibitem{mahboob_multimode_2014}
	\bibinfo{author}{Mahboob, I.}, \bibinfo{author}{Mounaix, M.},
	\bibinfo{author}{Nishiguchi, K.}, \bibinfo{author}{Fujiwara, A.} \&
	\bibinfo{author}{Yamaguchi, H.}
	\newblock \bibinfo{title}{A multimode electromechanical parametric resonator
		array}.
	\newblock \emph{\bibinfo{journal}{Scientific Reports}}
	\textbf{\bibinfo{volume}{4}}, \bibinfo{pages}{4448} (\bibinfo{year}{2014}).
	
	\bibitem{westra_nonlinear_2010}
	\bibinfo{author}{Westra, H. J.~R.}, \bibinfo{author}{Poot, M.},
	\bibinfo{author}{van~der Zant, H. S.~J.} \& \bibinfo{author}{Venstra, W.~J.}
	\newblock \bibinfo{title}{Nonlinear {Modal} {Interactions} in
		{Clamped}-{Clamped} {Mechanical} {Resonators}}.
	\newblock \emph{\bibinfo{journal}{Physical Review Letters}}
	\textbf{\bibinfo{volume}{105}}, \bibinfo{pages}{117205}
	(\bibinfo{year}{2010}).
	
	\bibitem{nayfeh_nonlinear_1979}
	\bibinfo{author}{Nayfeh, A.~H.} \& \bibinfo{author}{Mook, D.~T.}
	\newblock \emph{\bibinfo{title}{Nonlinear Oscillations}}
	(\bibinfo{publisher}{Wiley}, \bibinfo{year}{1979}).
	
	\bibitem{vogel_optically_2003}
	\bibinfo{author}{Vogel, M.}, \bibinfo{author}{Mooser, C.},
	\bibinfo{author}{Karrai, K.} \& \bibinfo{author}{Warburton, R.~J.}
	\newblock \bibinfo{title}{Optically tunable mechanics of microlevers}.
	\newblock \emph{\bibinfo{journal}{Applied Physics Letters}}
	\textbf{\bibinfo{volume}{83}}, \bibinfo{pages}{1337} (\bibinfo{year}{2003}).
	
	\bibitem{okamoto_coherent_2013}
	\bibinfo{author}{Okamoto, H.} \emph{et~al.}
	\newblock \bibinfo{title}{Coherent phonon manipulation in coupled mechanical
		resonators}.
	\newblock \emph{\bibinfo{journal}{Nature Physics}}
	\textbf{\bibinfo{volume}{9}}, \bibinfo{pages}{480} (\bibinfo{year}{2013}).
	
	\bibitem{seitner_classical_2016}
	\bibinfo{author}{Seitner, M.~J.} \emph{et~al.}
	\newblock \bibinfo{title}{Classical st\"uckelberg interferometry of a
		nanomechanical two-mode system}.
	\newblock \emph{\bibinfo{journal}{Physical Review B}}
	\textbf{\bibinfo{volume}{94}}, \bibinfo{pages}{245406}
	(\bibinfo{year}{2016}).
	
	\bibitem{fu_classical_2016}
	\bibinfo{author}{Fu, H.} \emph{et~al.}
	\newblock \bibinfo{title}{Classical analog of st\"uckelberg interferometry in a
		two-coupled-cantilever-based optomechanical system}.
	\newblock \emph{\bibinfo{journal}{Physical Review A}}
	\textbf{\bibinfo{volume}{94}}, \bibinfo{pages}{043855}
	(\bibinfo{year}{2016}).
	
	\bibitem{faust_nonadiabatic_2012}
	\bibinfo{author}{Faust, T.} \emph{et~al.}
	\newblock \bibinfo{title}{Nonadiabatic dynamics of two strongly coupled
		nanomechanical resonator modes}.
	\newblock \emph{\bibinfo{journal}{Physical Review Letters}}
	\textbf{\bibinfo{volume}{109}}, \bibinfo{pages}{037205}
	(\bibinfo{year}{2012}).
	
	\bibitem{unterreithmeier_damping_2010}
	\bibinfo{author}{Unterreithmeier, Q.~P.}, \bibinfo{author}{Faust, T.} \&
	\bibinfo{author}{Kotthaus, J.~P.}
	\newblock \bibinfo{title}{Damping of nanomechanical resonators}.
	\newblock \emph{\bibinfo{journal}{Physical Review Letters}}
	\textbf{\bibinfo{volume}{105}}, \bibinfo{pages}{027205}
	(\bibinfo{year}{2010}).
	
	\bibitem{ghadimi_dissipation_2016}
	\bibinfo{author}{Ghadimi, A.~H.}, \bibinfo{author}{Wilson, D.~J.} \&
	\bibinfo{author}{Kippenberg, T.~J.}
	\newblock \bibinfo{title}{Dissipation engineering of high-stress silicon
		nitride nanobeams}.
	\newblock \emph{\bibinfo{journal}{arXiv:1603.01605}}  (\bibinfo{year}{2016}).
	
	\bibitem{tsaturyan_ultra-coherent_2016}
	\bibinfo{author}{Tsaturyan, Y.}, \bibinfo{author}{Barg, A.},
	\bibinfo{author}{Polzik, E.~S.} \& \bibinfo{author}{Schliesser, A.}
	\newblock \bibinfo{title}{Ultra-coherent nanomechanical resonators via soft
		clamping and dissipation dilution}.
	\newblock \emph{\bibinfo{journal}{arXiv:1608.00937}}  (\bibinfo{year}{2016}).
\end{thebibliography}

\begin{thebibliography}{1}
	\expandafter\ifx\csname url\endcsname\relax
	\def\url#1{\texttt{#1}}\fi
	\expandafter\ifx\csname urlprefix\endcsname\relax\def\urlprefix{URL }\fi
	\providecommand{\bibinfo}[2]{#2}
	\providecommand{\eprint}[2][]{\url{#2}}
	
	\bibitem{pernpeintner_circuit_2014}
	\bibinfo{author}{Pernpeintner, M.} \emph{et~al.}
	\newblock \bibinfo{title}{Circuit electromechanics with a non-metallized
		nanobeam}.
	\newblock \emph{\bibinfo{journal}{Applied Physics Letters}}
	\textbf{\bibinfo{volume}{105}}, \bibinfo{pages}{123106}
	(\bibinfo{year}{2014}).
	
	\bibitem{saulson_thermal_1990}
	\bibinfo{author}{Saulson, P.~R.}
	\newblock \bibinfo{title}{Thermal noise in mechanical experiments}.
	\newblock \emph{\bibinfo{journal}{Physical Review D}}
	\textbf{\bibinfo{volume}{42}}, \bibinfo{pages}{2437--2445}
	(\bibinfo{year}{1990}).
	
	\bibitem{cleland_foundations_2003}
	\bibinfo{author}{Cleland, A.~N.}
	\newblock \emph{\bibinfo{title}{Foundations of Nanomechanics}}
	(\bibinfo{publisher}{Springer}, \bibinfo{year}{2003}).
	
	\bibitem{weaver_vibration_1990}
	\bibinfo{author}{Weaver, W., Jr.}, \bibinfo{author}{Timoshenko, S.~P.} \&
	\bibinfo{author}{Young, D.~H.}
	\newblock \emph{\bibinfo{title}{Vibration Problems in Engineering}}
	(\bibinfo{publisher}{Wiley}, \bibinfo{year}{1990}).
	
	\bibitem{nayfeh_nonlinear_1979}
	\bibinfo{author}{Nayfeh, A.~H.} \& \bibinfo{author}{Mook, D.~T.}
	\newblock \emph{\bibinfo{title}{Nonlinear Oscillations}}
	(\bibinfo{publisher}{Wiley}, \bibinfo{year}{1979}).
	
	\bibitem{unterreithmeier_coherent_2009}
	\bibinfo{author}{Unterreithmeier, Q.~P.}, \bibinfo{author}{Manus, S.} \&
	\bibinfo{author}{Kotthaus, J.~P.}
	\newblock \bibinfo{title}{Coherent detection of nonlinear nanomechanical motion
		using a stroboscopic downconversion technique}.
	\newblock \emph{\bibinfo{journal}{Applied Physics Letters}}
	\textbf{\bibinfo{volume}{94}}, \bibinfo{pages}{263104}
	(\bibinfo{year}{2009}).
	
	\bibitem{verbridge_high_2006}
	\bibinfo{author}{Verbridge, S.~S.}, \bibinfo{author}{Parpia, J.~M.},
	\bibinfo{author}{Reichenbach, R.~B.}, \bibinfo{author}{Bellan, L.~M.} \&
	\bibinfo{author}{Craighead, H.~G.}
	\newblock \bibinfo{title}{High quality factor resonance at room temperature
		with nanostrings under high tensile stress}.
	\newblock \emph{\bibinfo{journal}{Journal of Applied Physics}}
	\textbf{\bibinfo{volume}{99}}, \bibinfo{pages}{124304}
	(\bibinfo{year}{2006}).
	
	\bibitem{novotny_strong_2010}
	\bibinfo{author}{Novotny, L.}
	\newblock \bibinfo{title}{Strong coupling, energy splitting, and level
		crossings: {A} classical perspective}.
	\newblock \emph{\bibinfo{journal}{American Journal of Physics}}
	\textbf{\bibinfo{volume}{78}}, \bibinfo{pages}{1199} (\bibinfo{year}{2010}).
	
	\bibitem{seitner_classical_2016}
	\bibinfo{author}{Seitner, M.~J.} \emph{et~al.}
	\newblock \bibinfo{title}{Classical st\"uckelberg interferometry of a
		nanomechanical two-mode system}.
	\newblock \emph{\bibinfo{journal}{Physical Review B}}
	\textbf{\bibinfo{volume}{94}}, \bibinfo{pages}{245406}
	(\bibinfo{year}{2016}).
\end{thebibliography}
\end{document}